\documentclass{article}
\usepackage{spconf,amsmath,graphicx}
\usepackage{subcaption}

\newcommand\blfootnote[1]{%
  \begingroup
  \renewcommand\thefootnote{}\footnote{#1}%
  \addtocounter{footnote}{-1}%
  \endgroup
}

\title{VisionISP: Repurposing the Image Signal Processor for Computer~Vision~Applications}
\name{Chyuan-Tyng Wu$^{\ast}$, Leo F. Isikdogan$^{\ast}$\thanks{$^{\ast}$ Corresponding authors.}, Sushma Rao, Bhavin Nayak}
\secondlinename{Timo Gerasimow, Aleksandar Sutic, Liron Ain-kedem, Gilad Michael}
\address{Intel Corporation, Santa Clara, CA, USA}

\begin{document}
\ninept
\setlength{\textfloatsep}{10pt}
\maketitle

\begin{abstract}
Traditional image signal processors (ISPs) are primarily designed and optimized to improve the image quality perceived by humans. However, optimal perceptual image quality does not always translate into optimal performance for computer vision applications. We propose a set of methods, which we collectively call VisionISP, to repurpose the ISP for machine consumption. VisionISP significantly reduces data transmission needs by reducing the bit-depth and resolution while preserving the relevant information. The blocks in VisionISP are simple, content-aware, and trainable. Experimental results show that VisionISP boosts the performance of a subsequent computer vision system trained to detect objects in an autonomous driving setting. The results demonstrate the potential and the practicality of VisionISP for computer vision applications.
\end{abstract}
\section{Introduction}
\label{sec:intro}

Image Signal Processors (ISPs) carry out a series of processing steps to convert a raw signal acquired from an imaging sensor to a picture suitable for human consumption. Traditionally, ISPs have been tuned for optimal perceptual image quality to produce the best possible picture for human appreciation. Although humans are generally considered to be the primary consumers of images captured by cameras, machines have also started consuming images at a very large scale.

Following the resurgence of the deep artificial neural networks in the field of artificial intelligence, there has been an increasing research activity on high-level computer vision applications. Today, many of these applications run in parallel to a traditional ISP optimized for human viewers. However, such ISPs may not only fail to enhance images for better computer vision performance but may even degrade it. Therefore, many applications would benefit from an image processing pipeline that is optimized for computer vision applications.

The characteristics of input images that a computer vision system gets can differ greatly due to factors like different noise and optic profiles, noise level, and dynamic range. ISPs have this prior information that can explain some of the variations in the input. Indeed, a computer vision system can learn to compensate for those differences in the input signals. However, allowing the ISP to `normalize' the different physical characteristics of a camera system would help the computer vision system to use its representational capacity and processing power to perform higher-level tasks rather than to compensate for the low-level input variations. Additionally, ISPs usually have a substantial processing power for certain low-level operations such as denoising. This processing power would be fully utilized when an ISP is used to pre-process the data for computer vision applications.

\begin{figure}[tb]
\begin{center}
   \captionsetup[subfigure]{}
   \begin{subfigure}[t]{\linewidth}
   \begin{minipage}{\linewidth}
   \begin{center}
   \includegraphics[width=0.95\linewidth]{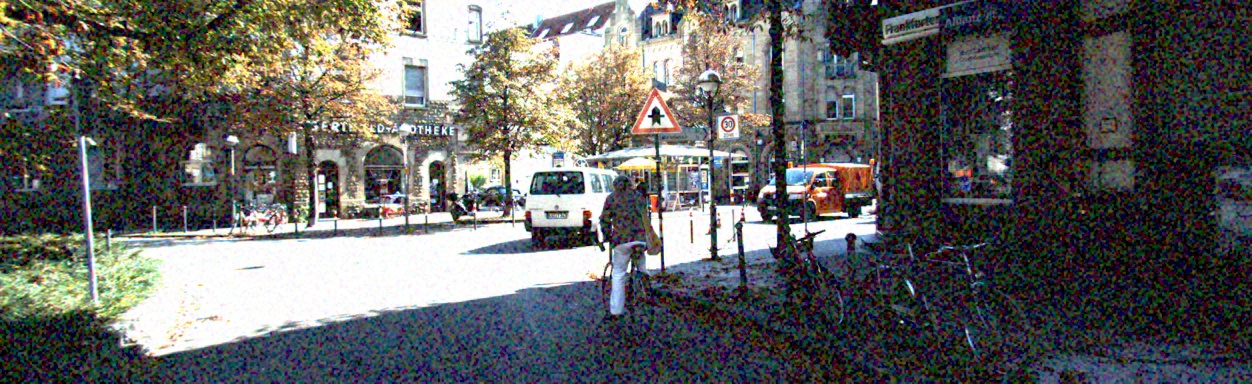}
   \end{center}
   \end{minipage}
   \vspace{-4pt}
   \caption{Input image}
   \end{subfigure}

   \medskip

   \begin{subfigure}[t]{\linewidth}
   \begin{minipage}{\linewidth}
   \begin{center}
   \includegraphics[width=0.95\linewidth]{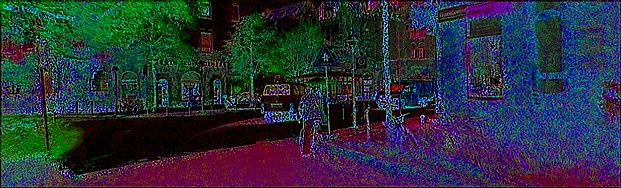}
   \end{center}
   \end{minipage}
   \caption{VisionISP output ($\times 2$ downscaling factor and $8\to 4$ bit-depth reduction)}
   \label{fig:tvs_exp_out}
   \end{subfigure}

\caption{Typical results delivered by VisionISP optimized for object detection. Although the output (\ref{fig:tvs_exp_out}) does not look natural to a human observer when visualized as a pseudo-color image, it `looks good' to a machine in the sense that it improves the object detection performance while using less bandwidth.}
\label{fig:tvs_example}
\end{center}
\end{figure}

\begin{figure*}[t]
\begin{center}
\includegraphics[width=1\linewidth]{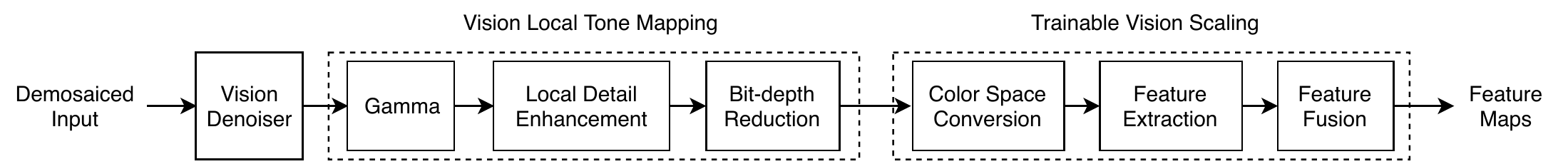}
\end{center}
\vspace{-10pt}
\caption{VisionISP consists of trainable blocks that repurpose an ISP for computer vision applications.}
\label{fig:vision_isp_flow}
\end{figure*}

For an imaging system designed with real-time computer vision applications in mind, the connection between an ISP and a computer vision engine would be as important as the processing blocks within these systems. It is a challenging problem to transmit high-resolution outputs with a large bit-depth and high frame rates from a sensor or an ISP to a computer vision engine given low power and low latency requirements. To reduce the data transmission load, it is common to downscale the input images or to lower the frame rate before they are fed into a computer vision engine. However, it is crucial for many applications, like autonomous driving, that neither the frame rate nor the relevant spatial information is reduced. Conventional downscaling methods, such as bilinear interpolation, are content agnostic. Therefore, small details in a scene, such as pedestrians, can easily be discarded during such downscaling despite their utmost importance. Maintaining high frame rates while preserving the relevant details would be possible through smart, content-aware operations.

In this work, we propose to repurpose the ISP for computer vision applications. First, we tune an existing ISP to find the optimal denoising configuration for a subsequent computer vision application. Then, we propose a sequence of operations that maintain high frame rates while preserving the details. These operations include tone mapping and feature-aware downscaling blocks that reduce both the number of bits per pixel and the number of pixels per frame. Our proposed operations are not only content-aware but also trainable. This processing sequence, which we call VisionISP, preserves the information that helps a subsequent neural network learn discriminative features even after significant data reduction. Our methods help reduce data transmission requirements, resulting in savings in power consumption and an overall better-optimized pipeline.

\vspace{-5pt}
\subsection{Related Work}
\label{sec:related_work}
Algorithms that constitute a typical imaging pipeline have been rigorously studied in the literature. Prior work on ISPs focused mostly on improving the image quality for human viewers~\cite{heide2014flexisp, autotune, hasinoff2016burst}. In recent work, the use of deep convolutional neural networks (CNNs) has become a common theme to improve image processing algorithms for a better imaging pipeline. Examples include models that perform demosaicing~\cite{Gharbi2016}, denoising~\cite{Mildenhall2018, ZhangZCMZ2017}, and many other types of image enhancement and transformation methods~\cite{ChenXK2017, chen2018learning, gharbi2017deep}.

Recently, there has also been interest in designing image restoration and enhancement methods with the overall goal of improving high-level computer vision performance rather than perceptual image quality~\cite{dirtypixels,Sharma2018}. Earlier work showed how image degradations affect image recognition accuracy~\cite{Dodge2016}, how conventional low-level image restoration techniques do not necessarily improve the computer vision performance~\cite{dirtypixels,buckler2017reconfiguring}, and how jointly handling image restoration and recognition tasks can improve the results~\cite{dirtypixels,Sharma2018,LiuWLWH2018}. It has also been shown that imaging devices can save power when they are used for computer vision tasks by running in a lower resolution and precision mode and skipping processing steps that are traditionally used for enhancing image quality~\cite{buckler2017reconfiguring}.

A major aspect that VisionISP differs from the prior work is that it repurposes parts of an existing ISP for computer vision applications without replacing them with proximals or large CNNs. Therefore, it can be integrated into low-power systems in a cost-effective way.

\vspace{-5pt}
\section{VisionISP Methodology}
VisionISP consists of blocks (Fig.~\ref{fig:vision_isp_flow}) that improve the input to a computer vision system while reducing the amount of data transmission. The first block is a computer-vision-driven denoiser that tunes an existing ISP without modifying the underlying algorithms and hardware design. The second and third blocks respectively reduce the bit-depth and resolution of their inputs while mitigating any negative impact on the computer vision performance.

\vspace{-5pt}
\subsection{Computer Vision Driven Image Denoising}
\label{sec:tisp}
The goal of a typical ISP is to produce clean and sharp images that improve the perceptual image quality for human viewers. Therefore, it is usually tuned based on photography-driven image quality characteristics that are important to the human vision system. However, tuning the ISP for the optimal image quality does not guarantee optimal results for computer vision tasks. For instance, heavier noise reduction might result in better perceptual image quality whereas it might be more beneficial for a machine vision system to tolerate a higher level of noise in exchange for more information.

We adopt the automatic ISP image quality tuning technique presented by Nishimura et al.~\cite{autotune} to optimize the denoising parameters of the imaging pipeline. Unlike the original automated ISP tuning procedure, we do not minimize the difference between a reference image and the ISP output to tune our pipeline. Instead, we minimize the content loss between those image pairs using a higher-level representation. We calculate this loss on feature maps extracted from a target neural network that is pre-trained to perform a particular computer vision task. In this way, we build a vision denoiser by tuning the denoiser for the computer vision task rather than perceptual image quality.

The vision denoiser operates on a demosaiced image. Indeed, the demosaicing block can be disabled to reduce the cost. However, we assume that the system already has an ISP running. Therefore the demosaicing step has no additional cost. Empirically, we found that the computer vision performance does not improve if we use a color filter array image rather than a demosaiced image.

\vspace{-5pt}
\subsection{Vision Local Tone Mapping (VLTM)}
Low-level details such as edges and corners are essential for many computer vision applications to be able to make accurate predictions. If we were to reduce the bit depth by some factor uniformly, this would result in the loss of those very details, thus degrading the performance of the following model. However, there is an excellent value in reducing the bit depth since it directly translates into simpler hardware (i.e., reduced area), reduced bandwidth, and significant savings in power. On some CNN accelerators, for example, running inference on 4-bit images can be twice as fast as compared to when using 8-bit images.

The goal of the vision local tone mapping (VLTM) operator is to reduce the bit-depth with a minimal impact on the accuracy, i.e., achieve similar accuracy with fewer bits per pixel. VLTM accomplishes this by using a global non-linear transformation followed by a local detail boosting operator before bit-depth reduction. The non-linear transform acts as a trainable global tone mapping operator while the detail boosting operator acts locally to preserve the details in the low-bit-per-pixel output. Both blocks operate on the luminance (Y) channel only.

\vspace{5pt} \noindent
{\bf Non-linear transformation}: We pass the luminance channel of an input image through a non-linear function to change its distribution before bit-depth reduction. This transformation preserves the portions of the dynamic range that are most relevant for the computer vision applications. This non-linear function can be any differentiable non-linear function having trainable parameters, such as a simple gamma function $Y' = Y_{in}^\gamma$.

\vspace{5pt} \noindent
{\bf Detail Enhancement}: We apply a simple box filter to $Y'$ to get a low-pass filtered signal $Y_\text{LPF}$. We use the difference between $Y'$ and the low pass filtered signal as a simple high-pass filter that accentuates the details. The output of this module uses a trainable parameter $\alpha$ to adjust the amount of detail as $Y_\text{out}=Y_\text{LPF}+\alpha\cdot (Y'-Y_\text{LPF})$.

\vspace{5pt} \noindent
{\bf Bit Depth Reduction}: We scale the bit-depth of the detail-boosted output signal $Y_\text{out}$ down to the desired number of bits by a simple bit-shift operation. The bit-depth reduction is performed after the first two operations to preserve the information that is relevant to the target computer vision task.
\vspace{5pt}

VLTM is computationally very cheap and can utilize existing hardware, such as a gamma correction module in an imaging pipeline. VLTM can be prepended to many types of neural networks to perform computer vision tasks more efficiently.

\subsection{Trainable Vision Scaler (TVS)}
\label{sec:tvs}
Trainable vision scaler (TVS) is a generic neural network framework that processes and downscales images before they are fed into a computer vision application. TVS consists of three main modules:

\vspace{5pt} \noindent
{\bf Color space conversion}: Many computer vision systems expect images or videos in RGB color space, whereas it is common for an ISP to use a different color space, such as YUV or YCbCr. To find an optimal color space automatically, we define a trainable color space conversion module instead of converting the input images into a particular color space. We define this color space conversion module as a simple $1\times 1\times 3 \times3$ convolutional layer with biases.

\vspace{5pt} \noindent
{\bf Feature extraction}: The feature extraction layer extracts low-level features that are important for computer vision tasks, such as edges and textures while reducing the input resolution. We define this module as a $K\times K\times 3\times N$ convolutional layer followed by a non-linear activation function, where N is the number of output filters, and $K$ is the filter size. This layer can also be defined as a combination of $K\times K\times 1\times N_1$, $K_1\times K_1\times 1\times N_1$, and $K_2\times K_2\times 1\times N_1$ filters to use different filter sizes to extract features at different scales. Empirically, we found that it suffices to use a single-scale convolutional layer of size $7\times 7\times 3\times 30$ for feature extraction. This module downscales its input up to a scaling factor of $K-1$ in a single pass, where the stride controls the scale factor. For example, a stride of two would downscale the input by a factor of two. Larger factors can be achieved via multiple passes through the TVS. The module also supports fractional scaling factors via flexible stride control. A non-integer stride is approximated using integer strides having different groups of sampling points. For instance, for a target scaling factor of 2.5, the sampling points would be the pixels at \{1.0, 3.5, 6.0, 8.5, 11.0, 13.5, 16.0, ...\}, which would become \{1, 4, 6, 9, 11, 14, 16, ...\} after rounding to the closest integers. Then, we can train a convolution kernel with a stride of 5, and sample the pixels at both \{1, 6, 11, 16, ...\} and \{4, 9, 14, 19, ...\}. This flexibility to have a non-integer scaling factor makes TVS practical for a wide variety of applications.

\vspace{5pt} \noindent
{\bf Feature fusion}: The feature fusion layer projects the feature maps produced by the feature extraction layer into three output channels since computer vision systems typically expect 3-channel (e.g., RGB) images as inputs. This layer can be defined as either a $1\times 1\times N\times 3$ convolutional layer or a group of three $1\times 1\times (N/3)\times 1$ convolutional layers, one per output channel. TVS outputs a set of three feature maps instead of a picture in a particular color space. We normalize these features to have zero-mean before they are fed into a computer vision system. The outputs of TVS can be visualized as pseudo-color images by mapping these feature maps into the R, G, B channels (Fig.~\ref{fig:tvs_exp_out}). Although such visualization may not look natural to human viewers, we show that it is a very efficient representation of what a camera should feed into a computer vision system to perform well (Sec.~\ref{sec:experiments_tvs}).

TVS is a very lightweight framework that can be implemented both in software and hardware. The computational cost of TVS can be further reduced by replacing the convolutional layer in the feature extractor with depthwise separable convolutions and quantizing the weights and activations.

\begin{table}[t]
\centering
\begin{tabular}{l||ccc}
\hline
Noise variance           & 0.06   & 0.12   & 0.18   \\ \hline
Vision denoiser disabled  & 0.349  & 0.212  & 0.137  \\
Vision denoiser enabled & 0.421  & 0.279  & 0.216  \\ \hline
\textbf{Relative improvement} & \textbf{20.6\%} & \textbf{31.6\%} & \textbf{57.7\%} \\ \hline
\end{tabular}
\caption{The relative improvement in object detection performance (mAP) when the vision denoiser is used.}
\label{table:tisp_exp_result}

\vspace{10pt}

\begin{tabular}{l||ccc}
\hline
Bit depth   &      8 &      4 &      2 \\ \hline
VLTM disabled   & 0.778  &  0.772 &  0.694 \\
VLTM enabled       & 0.801 &  0.798 &  0.791 \\ \hline
\textbf{Relative improvement} & \textbf{2.96\%} & \textbf{3.37\%} & \textbf{13.98\%} \\ \hline
\end{tabular}
\caption{The relative improvement in object detection performance (mAP) when the VLTM is used.}
\label{table:vltm_exp_result}
\end{table}

\vspace{-5pt}
\section{Experiments}
To evaluate VisionISP, we chose object detection as a representative target computer vision task, since it encompasses both recognition and localization aspects of computer vision tasks. We used the KITTI 2D object detection dataset~\cite{kitti} to train and validate the components of VisionISP. The KITTI dataset consists of driving footages having annotations for cars, pedestrians, and cyclists. We measured the impact of each component of VisionISP in terms of relative gains in mean average precision (mAP). We used the code provided by the authors of SqueezeDet~\cite{squeezedet} in our experiments using the default configurations for training. We plugged in VisionISP to the SqueezeDet framework with minimal modification in the original code.

\vspace{-5pt}
\subsection{Denoising}
We evaluated the VisionISP denoiser on pairs of noisy and denoised images. We generated the noisy images by injecting Gaussian noise to the images in the KITTI dataset in varying levels. We used the original clean images as a reference in the automatic ISP tuning step. We tuned the denoiser to minimize the sum of absolute differences between feature map activations extracted from the backbone neural network. Automatically tuning the ISP to perform computer vision driven image denoising led to a relative improvement of up to 57.7\% in the object detection precision (Table~\ref{table:tisp_exp_result}). The gains from denoising increased as the noise variance increased.

\vspace{-5pt}
\subsection{Vision Local Tone Mapping}
We evaluated VLTM by measuring its impact on object detection performance when the subsequent backbone network is trained on reduced bit-depth inputs. We chose the non-linear function in VLTM to be a gamma function. Therefore VLTM had only two trainable parameters $\alpha$ and $\gamma$. We trained VLTM and the backbone object detection network from scratch end-to-end.

As expected, VLTM offered greater value when the input is compressed into fewer bits (Table~\ref{table:vltm_exp_result}). In the experiments, VLTM was able to compress the bit depth from 8 bits to up to 2 bits without a significant hit on the object detection performance. Therefore, VLTM can be used to reduce power consumption by a factor of 4 while retaining the accuracy.

\begin{figure}[t]
\begin{center}
\includegraphics[width=0.98\linewidth]{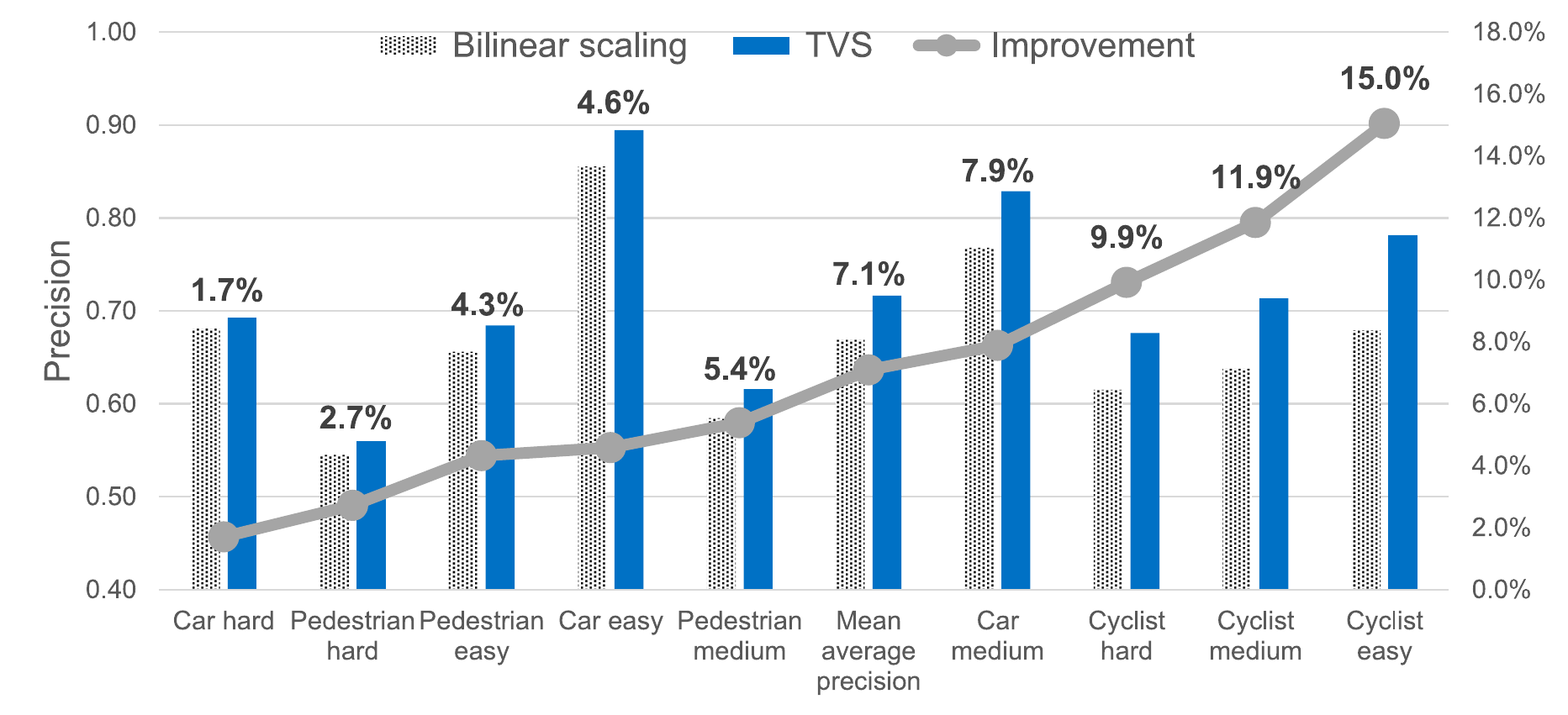}
\end{center}
\vspace{-20pt}
\caption{Relative improvements when TVS is used to downscale the input $\times 2$ as compared to bilinear scaling. TVS and SqueezeNet backbone are trained end-to-end.}
\label{fig:tvs_exp_sdp_scratch_2}
\end{figure}

\begin{figure}[t]
\begin{center}
\includegraphics[width=0.98\linewidth]{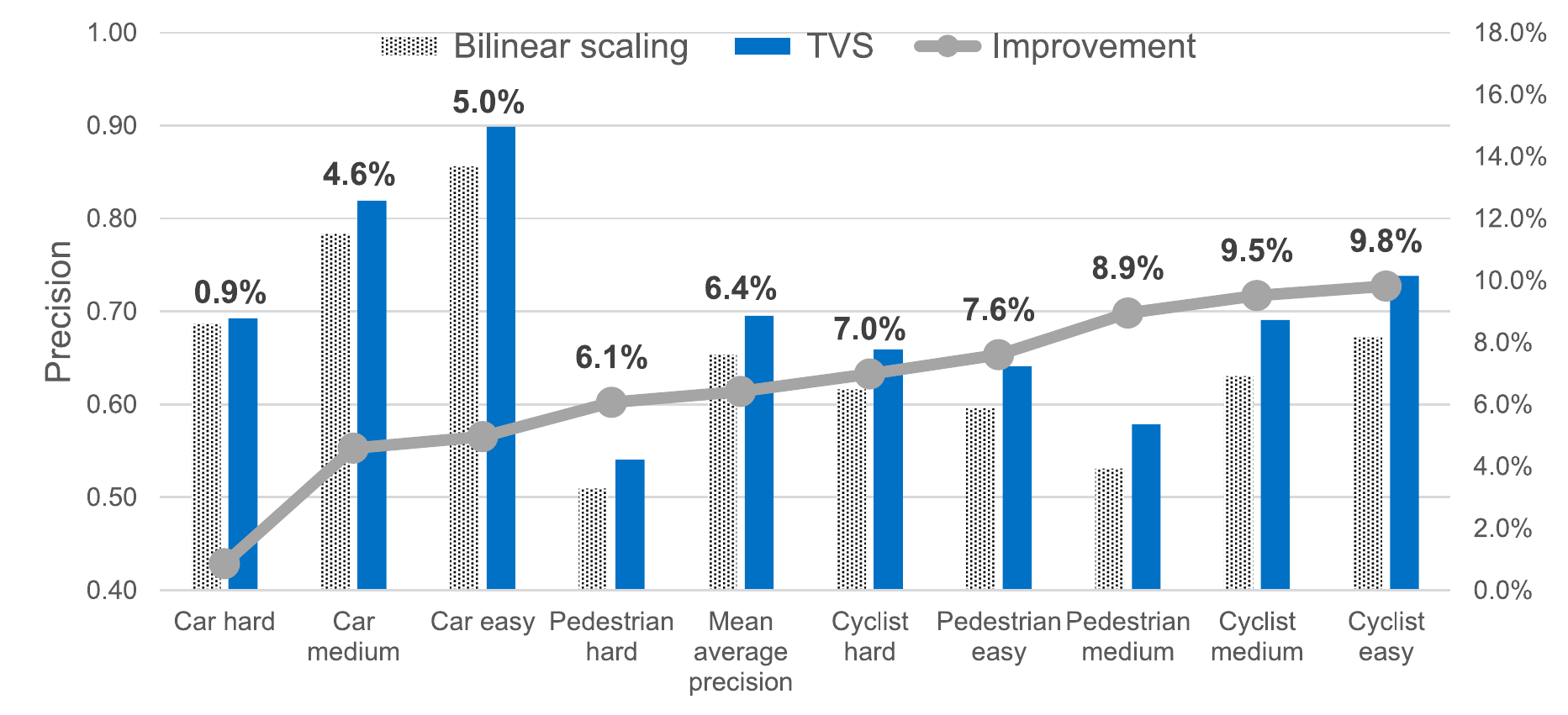}
\end{center}
\vspace{-20pt}
\caption{Generalizability: TVS is trained using a SqueezeNet backbone and evaluated on a ResNet50 backbone.}
\label{fig:tvs_exp_res_notrain_2}
\end{figure}

\begin{figure}[t]
\begin{center}
\includegraphics[width=0.98\linewidth]{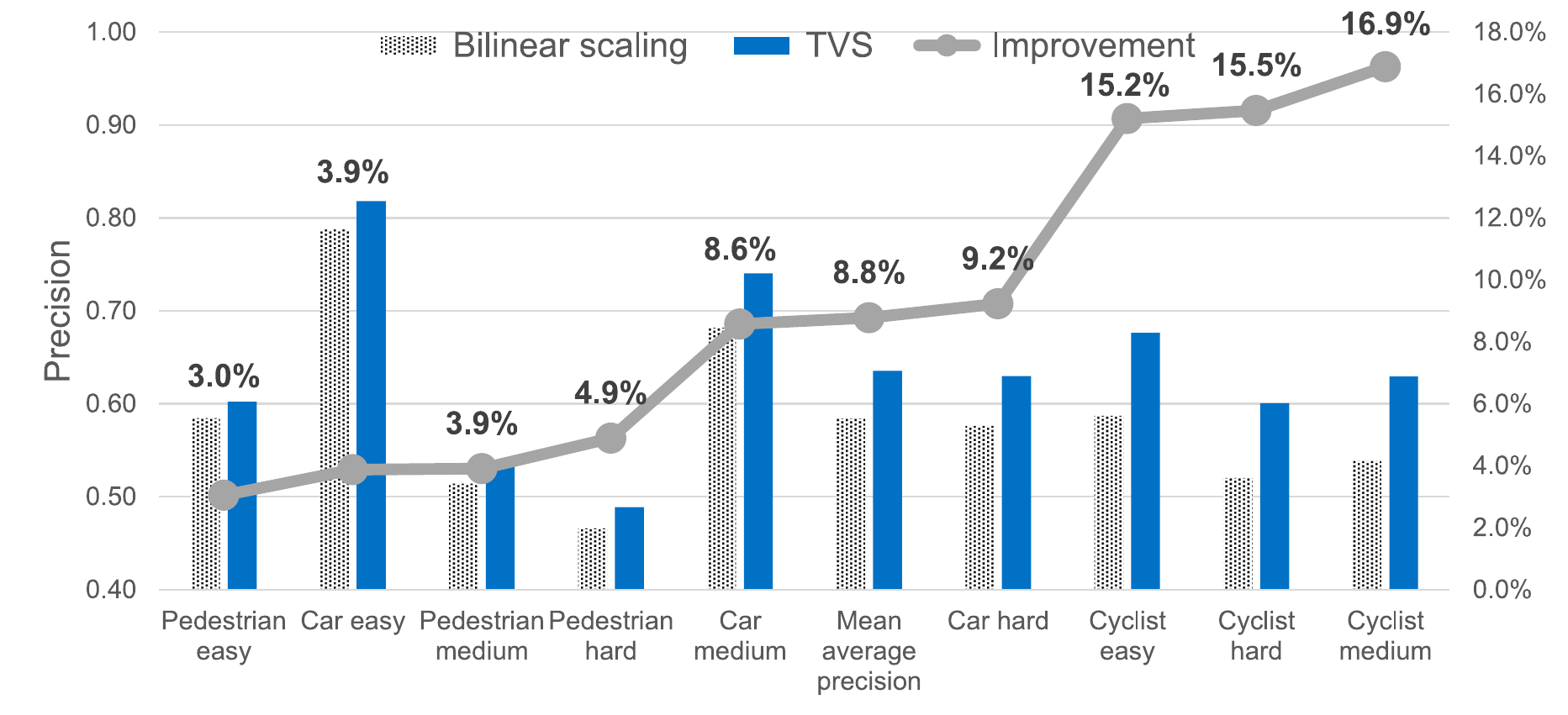}
\end{center}
\vspace{-20pt}
\caption{Scaling flexibility: TVS is trained with a downscaling factor of $\times 2$ and evaluated with a downscaling factor of $\times 2.5$.}
\label{fig:tvs_exp_sdp_notrain_2p5}
\end{figure}

\vspace{-5pt}
\subsection{Trainable Vision Scaler}
\label{sec:experiments_tvs}
We trained the object detection models where the input images are downscaled by two using TVS, using bilinear scaling as a baseline. In our first experiment, we trained the models using the SqueezeDet framework with a SqueezeNet~\cite{iandola2016squeezenet} backbone feature extractor. We trained TVS and the backbone network end-to-end. Using TVS has improved the object detection performance for all classes in the KITTI dataset (Fig.~\ref{fig:tvs_exp_sdp_scratch_2}) as compared to using bilinear interpolation with the same scaling factor.

We also evaluated the generalizability of TVS by training it for one backbone network and validating it in another without any further training or fine-tuning of the block. We first trained TVS with a downscaling factor of  $\times 2$ end-to-end with a SqueezeNet backbone. Then, we plugged the trained TVS into a ResNet50~\cite{ResNet} backbone and evaluated its performance, training only the backbone network while the weights of TVS stayed frozen (Fig.~\ref{fig:tvs_exp_res_notrain_2}). Using the same weights, we also changed the downscaling factor to $\times 2.5$ using flexible stride control (Sec.~\ref{sec:tvs}) and evaluated its performance using the SqueezeNet backbone to observe how well a trained TVS generalizes to slightly different scaling factors (Fig.~\ref{fig:tvs_exp_sdp_notrain_2p5}).

The results showed that once TVS is trained, it can be deployed to other computer vision systems as-is. Use of TVS improved the object detection performance as compared to bilinear interpolation, even when it is not optimized for the particular neural network and scaling factor that it was evaluated on. This generalization characteristic gives TVS the flexibility to be used in cases where the target computer vision system is not available for training.

\vspace{-5pt}
\subsection{Jointly Training VLTM and TVS}
We trained VLTM and TVS with a downscaling factor of two simultaneously with an object detection model on the KITTI dataset. We moved the bit-depth reduction to the end of TVS to co-utilize the blocks better. Although the gains from VLTM and TVS were not merely additive, jointly training VLTM and TVS provided further gains in the object detection performance. Enabling VLTM improved the object detection performance when used together with TVS, even when the input bit depth is reduced (Table~\ref{table:visp_exp_result}).

In the experiments, we had full access to both VisionISP and the backbone neural network performing the object detection. Therefore we were able to train VisionISP and the backbone computer vision model end-to-end jointly. However, VisionISP can also be trained separately and deployed as-is with frozen parameters. Even when the underlying model is not known or available, VisionISP would still provide value when it is trained using a known model with a similar target.

\begin{table}[t]
\centering
\begin{tabular}{l || c c c | c c}
\hline
Bit depth & $8\to 8$ & $8\to 8$ & $8\to 8$ & $8\to 4$ & $8\to 4$ \\
\hline
VLTM      & Bypass & Bypass & Enable & Bypass & Enable \\
TVS       & Bypass & Enable & Enable & Bypass & Enable \\
\hline
mAP       &  0.735 &  0.759 &  \textbf{0.782} &  0.729 &  \textbf{0.762} \\
\hline
\end{tabular}
\caption{The impact of using VLTM and TVS ($\times 2$ scaling) on object detection performance (mAP) when they are trained jointly. The models are trained end-to-end with a SqueezeNet backbone network initialized with ImageNet~\cite{imagenet} pre-trained weights.}
\label{table:visp_exp_result}
\vspace{5pt}
\end{table}

\vspace{-10pt}
\section{Conclusion}
We have presented a set of methods to improve the traditional ISP towards optimizing its performance for computer vision applications. Our proposed framework, VisionISP, consists of three primary processing blocks: vision denoiser, VLTM, and TVS. The denoiser reduces the noise in the input signal while preserving the image content for a computer vision system by modifying the tuning targets on an existing ISP. VLTM and TVS significantly reduce the amount of transmitted data between the ISP and a computer vision engine without a significant negative impact on the performance. The added value of VLTM and TVS is particularly large at higher levels of bandwidth compression. Each one of the blocks in VisionISP improves the performance of a subsequent computer vision system, therefore can be deployed both independently and in tandem with each other. We evaluated VisionISP for object detection using an autonomous driving benchmark dataset and VisionISP proved to have practical potential. We believe further research and development would help discover more possibilities in a wide range of computer vision applications.

\bibliographystyle{IEEEbib}

\blfootnote{This manuscript is the accepted version of the following paper: \\ \\
``VisionISP: Repurposing the Image Signal Processor for Computer Vision Applications," 2019 IEEE International Conference on Image Processing (ICIP), Taipei, Taiwan, 2019, pp. 4624-4628. doi: 10.1109/ICIP.2019.8803607. \\ \\
{\small\textcopyright} 2019 IEEE. Personal use of this material is permitted. Permission from IEEE must be obtained for all other uses, in any current or future media, including reprinting/republishing this material for advertising or promotional purposes, creating new collective works, for resale or redistribution to servers or lists, or reuse of any copyrighted component of this work in other works.\\}

\end{document}